# Do cell culturing influence the radiosensitizing effect of gold nanoparticles part 1: scrutinizing recent evidence for data consistency

Hans Rabus[1], Oswald Msosa Mkanda[1,2,3]

[1] Physikalisch-Technische Bundesanstalt, Berlin, Germany
[2] University Dar es Salaam, Dar es Salaam, Tanzania
[3] Mzuzu University, Mzuzu, Malawi

E-mail: hans.rabus@ptb.de

**Abstract**

In radiobiological experiments, the cells can either float in suspension or adhere to the walls of the sample holder. When irradiation is performed in the presence of dose modifying agents such as gold nanoparticles (AuNPs), the different shapes of the floating or adherent cells may imply a different dose to the nucleus, with biological consequences such as cell survival. Recently, it has been reported that the survival rate varies by up to a factor of 1.5 for the two cell geometries and by up to a factor of 2 for different orientation of the cells with respect to the incident beam. These results are examined in this paper and possible methodological issues are analyzed. This analysis shows that the simulation setup corresponds to the case of cells in the dose build-up region near the surface of a water phantom, where different depths result in different dose and survival probabilities. The validation of the simulations by comparison with experimental data is misleading, as the apparent agreement is due to a neglect of the quadratic term of the linear-quadratic survival model. The analysis further shows that the reported step-like changes between the survival predicted from the mean dose and the LEM could be explained by the fact that in the entire simulation only in one event an ionizing interaction of a photon took place in an AuNP. It is shown that the probability of this is in the permille range and that the total number of electrons leaving an AuNP, estimated from the reported electron spectrum, is three orders of magnitude higher than the value estimated from the expectation of photon interactions in the AuNPs. This contradiction would be resolved if the AuNP diameter in the simulations were a factor of 10 larger than intended. Another possible explanation for the discrepancies is a hidden bias in the simulation geometry, for example, if the distribution of AuNPs was non-uniform.

## 1. Introduction

When biological cells are irradiated in vitro, their geometry may depend on whether they are in suspension (i.e. freely floating in the surrounding medium) or adherent to the bottom or side walls of the Petri dish or the wells of a multi-well plate containing them and the surrounding medium. In suspension, one can expect a more spherical shape (Bilodeau, 2024) which minimizes surface tension. In an adherent cell, however, surface tension is minimized when the cell is flattened out.

When irradiated under charged particle equilibrium conditions, the dose to a microscopic volume is the same regardless of its shape if the material density and composition are the same as in its surroundings. This should also be true if the irradiated volume is in the dose build-up region when the center of gravity of the different shapes is at the same depth in the phantom, so that the irradiation conditions are the same.

When cells are irradiated in the presence of dose-modifying agents such as nanoparticles (NPs) made of high atomic number Z material, the dose to the nucleus and the resulting cell survival probability depend on the spatial distribution of the NPs in and around the cell (Sung et al., 2017). Therefore, a different shape of the cell may affect the cell survival probability when NPs are present, similar to the different radiosensitivity of adherent and suspended cells reported by (Cansolino et al., 2015).

A recent simulation study reported that the surviving fraction (SF) of cells containing gold NPs (AuNPs) strongly depends on cell morphology and on the irradiation direction and concludes that "the direction of irradiation plays a crucial role in determining the effectiveness of AuNPs in reducing SF" (Antunes et al., 2025). The underlying simulations were based on detailed voxelized models of the cells derived from confocal microscopy imaged (Antunes et al., 2024). A $^{60}$Co irradiation facility and the Bragg-peak region of a proton beam with an initial energy of 14 MeV were considered as irradiation conditions.

The results reported for $^{60}$Co irradiation are presented in Fig. 1, which shows the predicted reduction of the surviving fraction at a dose of 2 Gy according to the variant of the local effect model (LEM) used by (Antunes et al., 2025). Fig. 1(a) corresponds to a cell in suspension with a similar cross-section viewed from above (z-axis) and from the side. Fig. 1(b) shows the results for an adherent cell in which the two cross-sections differ by a factor of about 2 (Antunes et al., 2025).

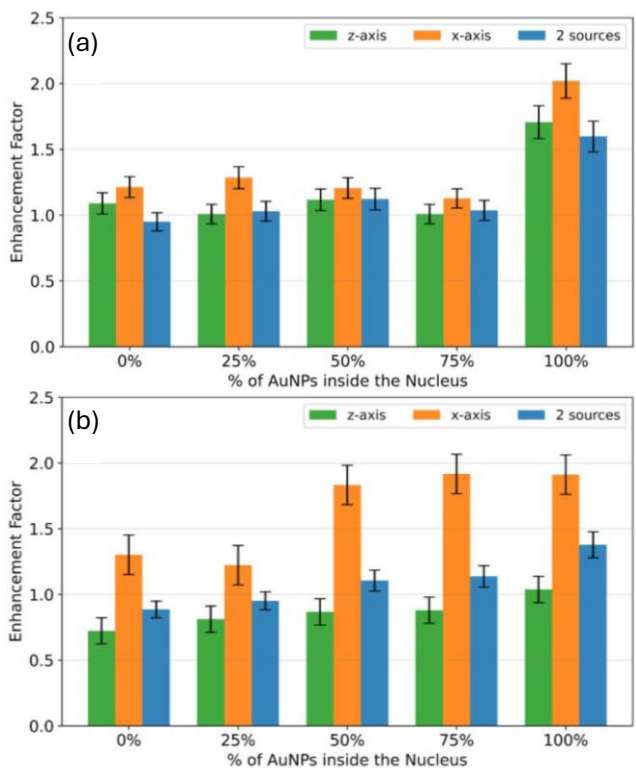

Fig. 1: Results for the predicted reduction of cell survival for (a) PC3 cells in suspension and (b) adherent PC3 cells when irradiated by $^{60}$Co gamma rays at a dose of 2 Gy, with 4326 spherical gold nanoparticles with a diameter of 4.79 nm present in the cell with different proportions in the cell nucleus. Reproduced from (Antunes et al., 2025) (copyright Antunes et al.) under the CC BY 4.0 license (http://creativecommons.org/licenses/by/4.0/). The panels have been rearranged to fit in one column, and the labels "A" and "B" replaced by "(a)" and "(b)".

The different groups of columns correspond to different proportions of the AuNPs in the cell nucleus, and the different colors correspond to irradiation from the top (z-axis), from the side (x-axis), and by two sources placed as in the irradiator facility. The y-axis of the plot shows the reduction in cell survival compared to cells without AuNPs exposed to the same photon fluence producing a dose of 2 Gy in the case without AuNPs. (In (Antunes et al., 2025), the ratio of survival without to survival with AuNPs is defined as the "survival enhancement factor".)

Fig. 1(a) shows that for a cell in suspensions, a reduced cell survival with AuNPs is predicted. With all AuNPs in the nucleus of a cell the survival is reduced by a factor between 1.5 and 2, depending on the irradiation geometry. The data for the adherent cell in Fig. 1(b) suggests that the cell survival is by a factor of about 2 different depending on whether the radiation impinges on the larger or on the smaller cross-section of the cell. For the irradiation along the z-axis, a higher survival probability is predicted in most cases when a cell contains AuNPs as compared to when there are no AuNPs. The exception is the case of all AuNPs located in the cell nucleus.

The predicted increase of survival with AuNPs present during irradiation appears highly implausible. Therefore, it may be suspected the results reported by (Antunes et al., 2025) for $^{60}$Co irradiation may be simulation artefacts. Simulation results for AuNPs may vary strongly depending on the parameters used (Vlastou et al., 2020), where the simulation geometry is particularly relevant for microscopic simulations of AuNPs (Zygmanski et al., 2013).

Leaving aside the case of proton irradiation for which similar issues exist, this first part of the paper scrutinizes the results of (Antunes et al., 2025) for cells irradiated at a $^{60}$Co facility and suggests explanations for the findings that conflict with basic radiation physics. The focus here is on problems that could be solved by additional simulations with a slight revision of the simulation setup and strategy. The second part of the paper deals with potential biases caused by the methodology used by (Antunes et al., 2025) that cannot be eliminated by additional simulations alone.

## 2. Materials and Methods

### 2.1 Simulation set-up for the $^{60}$Co cell irradiation

(Antunes et al., 2025) used a two-step simulation approach. In the first step, the irradiation of a multi-well plate (MWP) was simulated in the $^{60}$Co irradiation facility used for preceding cell experiments. The particles impinging on the MWP were recorded in a phase space file (PSF), which was

2/14

then resized to a cross section "matching that of the cell model" (Antunes et al., 2025). The transformed PSF was used as the radiation source in the second simulation by random sampling from its entries. In this simulation, the particle source was placed at a distance of 0.5 µm from the cells.

This geometry is equivalent to considering only cells at a distance of 0.5 µm from the surface of the MWP or, if the wells were not filled to the top, from the surface of the medium containing the cells in a well. Therefore, only cells at the very beginning of the dose buildup region were considered. This implies a potential bias of the results for different cell geometries and orientations, resulting from a different mean distance of the cell center from the source. In addition, the secondary particle component of the radiation field is much smaller than at greater depths.

To estimate the effects of placing cell models of different sizes and orientations near the radiation source on the results, the geometric dimensions of the cells were estimated using the information and plots provided in (Antunes et al., 2025). For this purpose, the ellipses and rectangles with blue dashed contours shown in Fig. 2 were visually fitted to the cross-sections of the cell models shown.

By calculating the corresponding areas in paper coordinates and comparing them with the values given by (Antunes et al., 2025), the linear dimensions of the large blue ellipse in Fig. 2(a) and the rectangle in Fig. 2(b) were obtained. By determining the center of the cells (indicated by the dot-dashed lines on the left) and a line at 0.5 µm from the cell model (dot-dashed lines on the right), the distances between the source and the center of the cells for irradiation along the $x$-axis were determined. The dashed lines to the left of the cells are also 0.5 µm from the cell, and the second dashed lines indicate the centers of the cell nuclei.

The thickness in the $z$-direction was estimated by counting the voxels in Supplementary Fig. 1 and multiplying by the length of the voxels in the $z$-direction.

## 2.2 Cell survival model

(Antunes et al., 2025) fitted the linear-quadratic (LQ) model for the dependence of survival fraction $S$ on dose $D$ (Eq. (1)) to their data of experimental survival fractions to obtain the LQ model parameters $\alpha$ and $\beta$.

$$S = e^{-\alpha D - \beta D^2} \tag{1}$$

These parameters were then used with the local effect model (LEM) to determine cell survival from the simulations with AuNPs. For this purpose, the local dose and square of the dose in a voxelized geometry of the cell nucleus were evaluated. From these values per voxel, the mean number of lethal lesions which determine the survival of the cell is obtained by averaging over the cell nucleus. The resulting cell survival probability is given by Eq. (2):

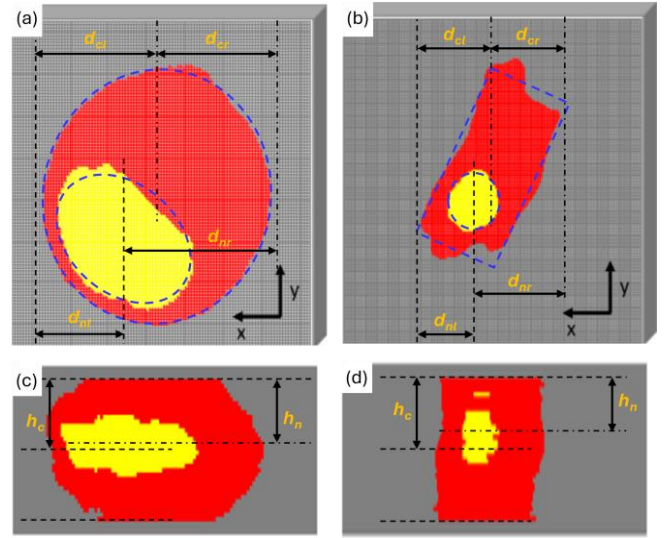

Fig. 2: Cross-sections in the $x$-$y$-plane of the models for (a) a cell in suspension and (b) an adherent cell. (c) and (d) show the corresponding cross-sections in the $y$-$z$-plane. Reproduced from (Antunes et al., 2025) (copyright Antunes et al.) under the CC BY 4.0 license (http://creativecommons.org/licenses/by/4.0/) with the following modifications: The panels were cut out, rearranged and newly labeled, and dot-dashed and dashed lines, double-arrows and their labels were added. In addition, the pictures in (c) and (d) were scaled in both directions to have the same height and a width matching that of the pictures in (a) and (b).

$$S_{\text{LEM}} = e^{-\alpha \overline{D} - \beta \overline{D^2}} \tag{2}$$

In Eq. (2), $\overline{D}$ and $\overline{D^2}$ are mean values over the $n$ voxels in the cell nucleus according to Eq. (3), where $D_i$ and $d_i$ denote the dose in voxel $i$ and the corresponding dose per event in the second simulation, respectively.

$$\overline{D} = \frac{1}{n}\sum_{i=1}^{n} D_i \;\;;\;\; \overline{D^2} = \frac{1}{n}\sum_{i=1}^{n} D_i^2 \;\;;\;\; D_i = N d_i \tag{3}$$

For the simulations without AuNPs, the mean dose was used to obtain the survival prediction.

$$S_{\text{MD}} = e^{-\alpha \overline{D} - \beta \overline{D}^2} \tag{4}$$

In the paper of (Antunes et al., 2025), the survival probabilities for irradiation with AuNPs according to Eq. (2) is compared with that according to Eq. (4). From the ratio between the two survival predictions, it is possible to estimate the value of $\overline{D^2}$ obtained by (Antunes et al., 2025) by using Eq. (5):

$$\overline{D^2} = \frac{1}{\beta}\ln\left(\frac{S_{\text{MD}}}{S_{\text{LEM}}}\right) + \overline{D}^2 \tag{5}$$

Here ln denotes the natural logarithm.

## 2.3 Number of events with electrons leaving the AuNP

In the work of (Antunes et al., 2025), only the normalized fluence of electrons leaving the AuNP was shown and is reproduced for the case of $^{60}$Co irradiation in Supplementary Fig. 3. Estimates of the absolute values were obtained in the following way: The width of the energy interval was estimated by counting the number of histogram bars per energy interval. The graphs were imported into Excel and rectangular shapes were superimposed to measure the height of the smallest histogram bar (indicated by the arrow in Supplementary Fig. 3) and the tick interval on the $y$-axis. The ratio of the former to the latter was used to determine the lowest value of the relative spectral fluence in the simulations. Assuming that the smallest value represents the case in which only one count in this energy interval was obtained in the simulation, the total electron count in the simulation results was estimated by the ratio of the area under the histogram to the area of the bar in the bin with the lowest fluence.

## 2.4 Number of ionizing interactions in an AuNP

The estimated number $n_i$ of ionizing interactions of photons in an AuNP was determined according to a further development of the approach presented in (Rabus et al., 2019). In this approach, generalized spline interpolations of the mass attenuation coefficient $(\mu_i/\rho)_g$ for ionizing interaction in gold are used as a weighting factor for the photon fluence. The coefficients of the spline functions can be found in (Rabus et al., 2019). In the present development, a generalized spline function for the attenuation coefficient of coherent scattering in gold was determined and subtracted from the total interaction coefficient to determine only interactions that lead to ionization in the AuNP.

The ratio of the expected number $n_i$ of ionizing photon interactions in an AuNP to the photon fluence $\Phi_p$ is given by Eq. (6):

$$\frac{n_i}{\Phi_p} = \int \left(\frac{\mu_i}{\rho}\right)_g \Phi_{rel}(E) dE \times \rho_g V_{np} \quad (6)$$

Here, $\Phi_p$ is the total photon fluence at the AuNP, $\Phi_{rel}(E)$ is the relative spectral fluence (normalized to unity), and $\rho_g$ and $V_{np}$ are the density of gold and volume of the AuNP, respectively. For numerical evaluation of the integral, the right-hand side was replaced by a sum over the energy bins, using the photon energy spectrum of the $^{60}$Co calibration facility at the German National Metrology Institute (PTB) shown in Supplementary Fig. 2.

With the results from Eq. (6), the expected number $n_{i,1}$ of ionizing interactions in an AuNP per incident photon can be determined by Eq. (7):

$$n_{i,1} = \frac{n_i}{\Phi_p} \times \frac{1}{A_{np}} \quad (7)$$

In the simulations of Antunes et al, a number $N_s$ of photons were emitted from the source with the area $A_s$. The corresponding fluence of an extended source is $N_s/A_s$. The expectation for the number of photons hitting one of $N_{np}$ AuNPs is then given by Eq. (8):

$$n_p = N_{np} \times A_{np} \times \frac{N_s}{A_s} \times \frac{\Phi_p}{\Phi_0} \quad (8)$$

Here, $\Phi_p/\Phi_0$ is the ratio of the photon fluence in the cell to the incident fluence. If the photon fluence $\Phi_p$ is replaced by the electron fluence $\Phi_e$ in Eq. (8), the corresponding expected number $n_e$ of electrons that hit an AuNP in the simulation is obtained. Eq. (9) provides the corresponding expression for the expected number of ionizing photon interactions in AuNPs in the simulations:

$$n_i = n_p \times n_{i,1} = \frac{n_i}{\Phi_p} \times N_{np} \times \frac{N_s}{A_s} \times \frac{\Phi_p}{\Phi_0} \quad (9)$$

(Antunes et al., 2025) multiplied the outcome of the second simulation, i.e. the dose per event sampled from the PSF, by the number $N$ of events calculated according to Eq. (10) to obtain the dose per decay of the $^{60}$Co source.

$$N_D = \frac{S_{cell}}{S_{plate}} \times t_D \times A \quad (10)$$

Where $S_{cell}$ is the cross-section of the cell (equal to $A_s$), $S_{plate}$ is the surface area of the MWP, $t$ is the irradiation time, and $A$ is the activity of the $^{60}$Co source. The first factor on the right-hand side of Eq. (10) compensates for the increase of particle fluence caused by the transformation of the PSF. The irradiation time is related to the dose rate $\dot{D}$ at the $^{60}$Co irradiation facility and the dose $D$ to be applied in the experiments by $t_D = D/\dot{D}$. The values obtained with Eq. (10) are used in this work to estimate the results per event in the simulations.

## 2.5 Simulations of the dose build-up in a $^{60}$Co field

The depth-dose curve in water for 60Co irradiation was simulated using the Geant4 toolkit(Agostinelli et al., 2003; Allison et al., 2006, 2016). A cubic water phantom of 400 mm side length was constructed using G4_WATER as material, housed in a surrounding air-filled cubic world volume of 3 m side length. In the water phantom, 400 slabs with a thickness of 1 mm and a cross-section of 350 mm × 350 mm were placed along the beam axis (z-direction) to serve as scoring volumes.

The simulation was performed using the Geant4 toolkit (version 11.2.2) employing the generic QBBC physics list, which includes standard electromagnetic processes, neutron tracking, and appropriate hadronic models in the relevant energy ranges, providing a balance between accuracy and computational effort. A global production cut of 0.1 mm was applied for photons, electrons, and positrons to define the energy range for secondary particle generation.

A general particle source (GPS) was positioned 100 cm from the surface of the phantom and employed to generate a conical beam of primary particles with a half aperture angle of 5°. The energy spectrum corresponded to that of the PTB 60Co calibration facility (see Supplementary Fig. 2). A total of $1\times10^8$ primary events were simulated.

Particle transport and energy deposition were scored using a customized G4UserSteppingAction class that captured physical quantities such as kinetic energy, step length, and energy deposition at every step within the scoring regions (slabs). The energy depositions within each slab were tallied and the total energy imparted was computed, enabling the determination of percentage depth dose (PDD) curves and other metrics. Data for all the quantities of interest was recorded in a .txt file for further processing. The PDD curve was determined based on the ratio of the dose per slab to that in the slab with the highest dose.

### 2.6 Dose around a 5 nm AuNP undergoing an ionization

In previous work (Thomas *et al* 2024), the radial dependence of the ionization clusters and energy imparted was determined, *inter alia* for a spherical AuNP with a diameter of 5 nm, which appears to be sufficiently close to the diameter of 4.89 nm used by (Antunes et al., 2025).

The data were generated in a two-step simulation. In the first step, the photon and electron fluence were determined in a scoring cylinder of 100 µm length in a water phantom with Geant4-DNA (Version 11.1.1) (Incerti et al., 2010a, 2010b, 2018; Bernal et al., 2015; Sakata et al., 2019). The center of the cylinder was 100µm away from the phantom surface. The cylinder had a radius of 10 cm, and the surrounding cylindrical phantom had a height of 100 cm and a radius of 50 cm to ensure a saturated contribution of backscattered photons. The simulations were performed with Geant4-DNA option 2, using track structure mode in the scoring cylinder and within a concentric cylindrical region with outer surfaces at a distance of 50 µm from those of the cylinder. Electron transport was suppressed outside this region. The incident radiation was a parallel photon beam with a radius of 10 cm (Thomas et al., 2024).

In the second simulation, the photon and electron fluence spectra from the first simulation were used for the source under the assumption of isotropic and laterally uniform irradiation of an AuNP. The electron transport was simulated in track structure mode in the AuNP and the surrounding water. In the simulations from which the data used here originate (Rabus and Thomas, 2025), the histories of primary particles leaving the AuNP were terminated, so that only the energy imparted from secondary particles was scored in spherical shells with a thickness of 5 nm starting from the AuNP surface. The contributions resulting from photons and electrons impinging on the AuNP were added using the ratios of the particle fluences per primary fluence from the first simulations as weights.

From the data, approximate estimates of the mean dose and mean square of the dose per voxel in the detailed cell model of (Antunes et al., 2025) were obtained by calculating the weighted averages of the two quantities over the radial shells around the AuNP falling within a sphere of radius $r_s$ according to Eq. (11):

$$r_s = \left(\frac{3}{4\pi}V_{\text{voxel}}\right)^{\frac{1}{3}} \tag{11}$$

This means that the sphere had same volume $V_{\text{voxel}}$ as the voxels.

## 3. Results

### 3.1 Consistence of the results of Antunes et al

The plots of the results of (Antunes et al., 2025) for irradiation without AuNPs are reproduced here as Fig. 3 with some additions made to highlight trends in the data. Fig. 3 show comparisons of the survival curves predicted from the simulation results with the cell survival data and the curve obtained with the best fit of Eq. (1) to the measured data. As can be seen from the dashed lines in Fig. 3, the predicted survival curves follow a linear trend in the semilogarithmic plots of survival fraction as a function of dose.

The predicted survival probability for the cell in suspension at a dose of 2 Gy without NPs is about 0.4 (blue arrow in Fig. 3(a)). The adherent cell shows similar survival probabilities with slight differences depending on cell orientation (Fig. 3(b)).

The survival probabilities at a dose of 2 Gy predicted for cells with AuNPs are shown in Fig. 4(a) for irradiation along the *z*-axis and in Fig. 4(b) for irradiation along the *x*-axis. The different groups of bars correspond to different ratios of the number of AuNPs in the nucleus to that in the cell. The filled bars are the predictions from the LEM (Eqs. (2) and (3)), the stroked bars show the prediction according to Eq. (4).

The horizontal dotted line in Fig. 4(a) shows the survival probability prediction for the adherent cell when irradiated without AuNPs along the *z*-axis. The corresponding value was read from the horizontal arrow in Fig. 3(a). The horizontal dotted line in Fig. 4(b) indicates a survival rate of 0.4, corresponding to that of the cell in suspension without AuNP indicated by the horizontal arrow in Fig. 3(a).

The long-dashed horizontal lines in Fig. 4(a) and Fig. 4(b) indicate the average height of the orange and green stroked bars, respectively. The former corresponds to the survival rate predicted from the mean dose for adherent cells irradiated along the z-axis, and the latter to the prediction for cells in suspension irradiated along the x-axis. For the cells in suspension, the survival rate predicted from the mean dose is about 0.5, which means that the mean dose with AuNPs was only about 1.7 Gy (cf. red dotted line in Fig. 3) when the cells in suspension were irradiated along the x-axis. The adherent cells show a dose reduction by a factor of about 0.7 when irradiated along the z-axis. A closer look at Fig. 3(b) shows that the survival prediction for the adherent cell is about 0.35 when irradiated along the x-axis. With the exception of the case where 100 % of the AuNPs are in the nucleus, the green bars in Fig. 4(b) are consistently above the dotted line that indicates a survival prediction of 0.4. Therefore, a dose reduction is also predicted in this case which amounts to a factor of about 0.8.

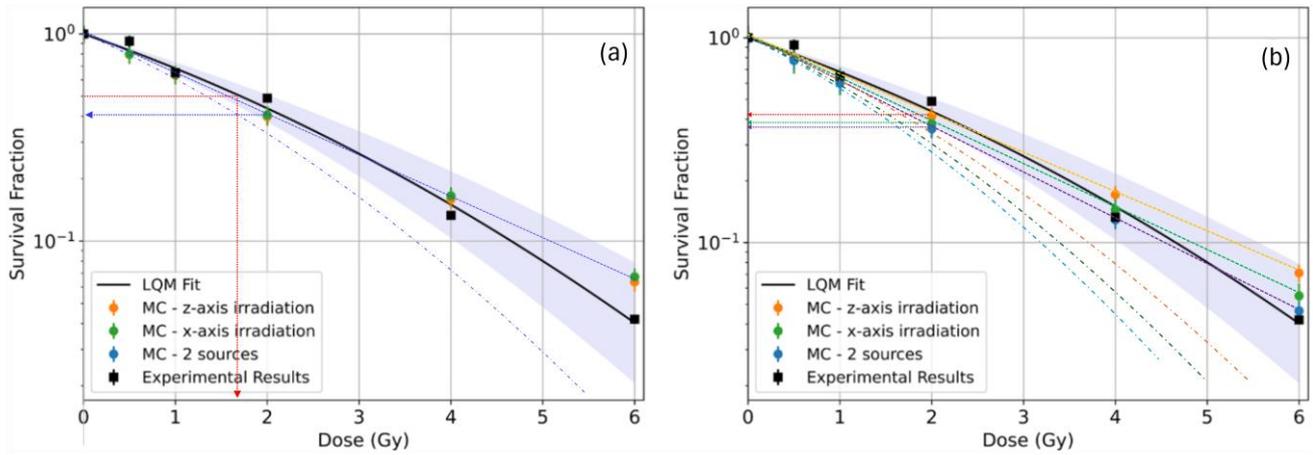

Fig. 3: Results for the predicted survival (colored symbols) of (a) PC3 cells in suspension and (b) adherent PC3 cells and experimental data (black symbols) from irradiation at a $^{60}$Co gamma source performed without GNPs. The solid black line is the fit curve to the experimental data according to Eq. (1). The dashed straight lines connect the colored datapoints; the dot-dashed lines represent the survival curve from the simulations under the assumption that the term with the square of the dose was omitted in the calculation leading to the dashed lines. The arrows pointing left indicate the predicted survival at a dose of 2 Gy; the red dotted line in (a) and the arrow pointing down indicate the dose corresponding to a cell survival of 0.5. Reproduced from (Antunes et al., 2025) (copyright Antunes et al.) under the CC BY 4.0 license (http://creativecommons.org/licenses/by/4.0/), modified by adding the dashed, dot-dashed, and dotted lines and arrows.

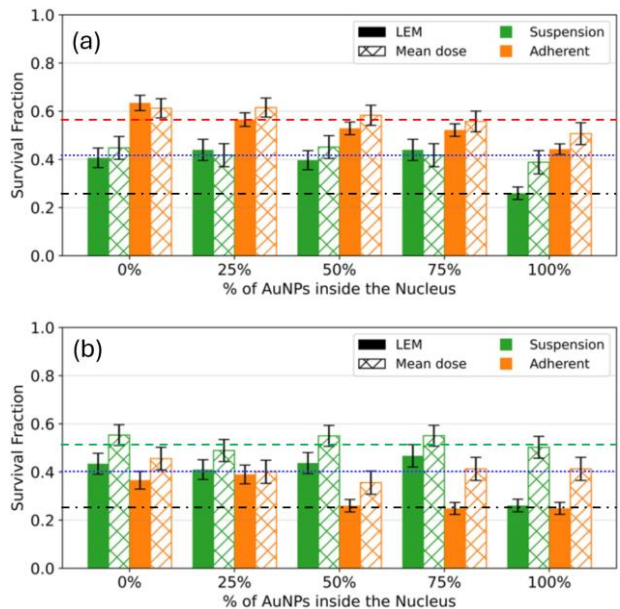

Fig. 4: Results for the survival fraction of PC3 cells containing 4326 spherical gold nanoparticles of 4.79 nm diameter at a dose of 2 Gy when irradiated along (a) the z-axis and (b) the x-axis by the particle spectrum from a $^{60}$Co gamma source. Reproduced from (Antunes et al., 2025) (copyright Antunes et al.) under the CC BY 4.0 license (http://creativecommons.org/licenses/by/4.0/), with the following modifications: cropping the top panel, changing the labels of the panels, adding the horizontal dashed, dotted and dot-dashed lines.

The horizontal dot-dashed lines in Fig. 4 indicate the lowest survival prediction in the two panels. This prediction corresponds to a survival rate of about 0.25. In general, it can be observed that the survival predictions shown in Fig. 4 tend to be close to one of the three horizontal lines.

For the case of cells in suspension and 100 % uptake into the nuclei, the survival prediction from the mean dose is the same as for a dose of 2 Gy without AuNPs, namely 0.4, while the prediction of the LEM is about 0.25. For the adherent cell, the survival rate with and without AuNPs at a dose of 2 Gy and irradiation along the x-direction is also about 0.40. For the cases where 75 % and 100 % of the AuNPs are in the nucleus, the predicted survival according to the LEM is also 0.25. In all cases, the estimated value for the mean square of the dose in the nucleus according to Eq. (5) is about 20 Gy².

### 3.2 Dose build-up

The cell dimensions estimated from Fig. 2 are listed in Table 1 together with the estimated distances between the

6/14

Table 1: Estimated dimensions of the cell models and distances between the center of the nucleus and the source in the simulations of (Antunes et al., 2025).

| parameter | suspension | adherent |
|---|---|---|
| long axis or side in *x-y*-plane / µm | 18.4 | 26.0 |
| short axis or side in *x-y*-plane / µm | 16.2 | 12.7 |
| thickness in *z*-direction / µm | 19.2 | 15.5 |
| source distances for *x*-irradiation | | |
| $d_{cr}$ / µm | 8.6 | 10.9 |
| $d_{cl}$ / µm | 8.7 | 10.8 |
| $d_{nl}$ / µm | 6.4 | 8.1 |
| $d_{nr}$ / µm | 10.9 | 13.3 |
| source distance for *z*-irradiation | | |
| distance to cell center / µm | 10.1 | 8.3 |
| distance to nucleus center / µm | 8.7 | 6.3 |

source plane and the centers of the cells and nuclei. The estimated distances of the source from the center of the cells for the two irradiation directions are about the same for the cell in suspension, whereas they differ by a factor of about 2 for the adherent. This indicates that there may be a bias, as different positions in the dose build-up region imply a different average dose to the cells.

Fig. 5 shows the depth-dose curve for the irradiation of a water phantom by a photon beam with an energy spectrum shown in the Supplementary Fig. 2. The symbols and dashed line show the average dose in 1 mm thick slabs of the phantom normalized to the maximum occurring value. The blue line is a fit to the datapoints in the dose-buildup region and shows an initial slope of about 70 % / mm.

### 3.3 Interactions in the simulations of Antunes et al

From Supplementary Fig. 3 and the methodology described in Section 2.3, the bin width was established as 10 keV and the ratio of total area under the histogram to the area of the histogram bar marked by the red arrow in Supplementary Fig. 3 is about 3900. Assuming that there is exactly one event contributing to this bin content, the total number of electrons scored is therefore also 3900. The ratio of the area of the first two energy bins of the histogram to that marked by the arrow is about 29. This energy range corresponds to that of the dominant Auger electrons of gold after core-shell excitation.

From the data obtained by Topas in (Rabus et al., 2021), it can be expected that about 4 or more Auger electrons are emitted from the AuNP when either the K-shell or one of the L-shells of a gold atom is ionized. Therefore, the number of ionizing photon interactions in an AuNP in the simulation cannot exceed 7. Since the lowest electron energy bins also contain low-energy secondary electrons, the number of ionizing photon interactions can be significantly smaller than this value.

With the photon spectrum shown in Supplementary Fig. 2, the expected number $n_{i,1}$ of ionizing interactions in an AuNP per incident photon according to Eq. (7) is $1.13 \times 10^{-6}$. Using this value, the expected number of ionizing interactions in AuNPs occurring in the simulations was calculated. The results obtained for 4326 AuNPs in the cells during irradiation and $10^7$ primary particles are shown in Table 2 together with those for the expected number of photons hits on an AuNP. The last column shows the ratio of the number $N_{2\,Gy}$ of events for a dose of 2 Gy according to Eq. (10) to the number $N_s$ of events in the simulations of (Antunes et al., 2025). The number $N_{2\,Gy}$ according to Eq. (3) was used by them to scale the results per event obtained from the simulations to a dose without AuNPs of 2 Gy.

Table 2: Cross sectional area $A_s$ of the photon source for the two cells and irradiation geometries (Antunes et al., 2025), expected number $n_p$ of events in which a photon hits a GNP (Eq. (8)), and expected number $n_i$ of events in which an ionization occurs in a GNP (Eq. (9)). The values apply to a simulation with $N_s = 10^7$ sampled events. The last column gives the ratio of the number of events $N_{2\,Gy}$ (corresponding to a dose of 2 Gy according to Eq. (10)) to $N_s$.

| case | $A_s$ / µm² | $n_p$ | $n_i$ | $N_{2\,Gy}/N_s$ |
|---|---|---|---|---|
| suspension | 240 | 3389 | $3.9 \times 10^{-3}$ | 30 |
| adherent, *z*-axis | 330 | 2465 | $2.8 \times 10^{-3}$ | 41 |
| adherent, *x*-axis | 167 | 4870 | $5.6 \times 10^{-3}$ | 21 |

### 3.4 Estimate for LEM effect from one responding AuNP

The local dose distribution around an AuNP of diameter 5 nm, in which an ionization occurs, is shown in Fig. 6. The black line shows the mean additional dose in spherical shells of 5 nm thickness resulting from ionization in the AuNP. The data are from (Rabus and Thomas, 2025) and correspond to an AuNP irradiated with the mixed radiation field at a depth of 100 µm in an extended water phantom resulting from irradiation with the 100 kV x-ray spectrum of (Thomas et al., 2024). The red line in Fig. 6 shows the average dose in a spherical shell around the AuNP with an outer radius corresponding to the value of the *x*-axis.

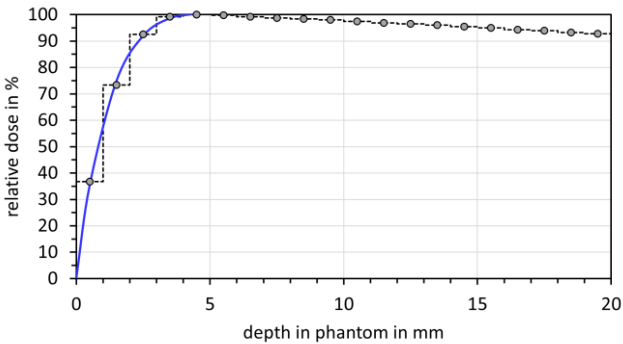

Fig. 5: Depth dose curve in a water phantom for the ⁶⁰Co photon energy spectrum shown in Supplementary Fig. 2. The blue line represents a fit to the first four datapoints.

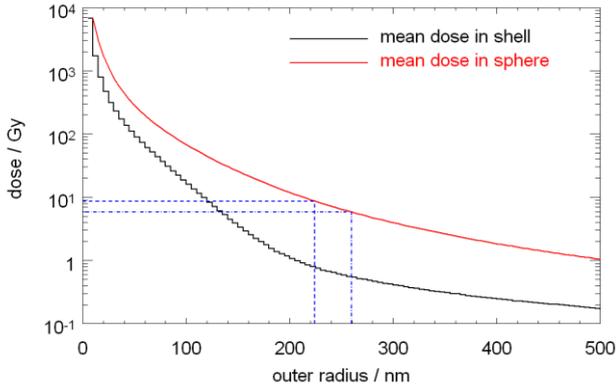

Fig. 6: Local dose in spherical shells around a GNP of diameter of 5 nm and average dose in a sphere around the GNP. The data correspond to the mixed photon and electron field at 100 μm depth in water resulting from an incident 100 kV x-ray spectrum (Rabus and Thomas, 2025). The dashed and dash-dotted vertical lines indicate the radii of spheres having the same volume as the voxels of the adherent and floating cell models, respectively. The horizontal lines indicate the corresponding mean dose in the voxel.

Table 3 shows the nucleus and voxel volumes for the adherent and suspended cells, the resulting number of voxels per nucleus and the radius of a sphere of the same volume as a voxel. The last line shows the mean dose in such a sphere when an AuNP located at the center undergoes an ionizing interaction. The corresponding radii of the spheres are shown in Fig. 6 by vertical lines and the resulting dose values by the horizontal lines. The dose value is 14.7 Gy for the adherent cell and 17.1 Gy for the cell in suspension.

In the simulations, the scored value of the dose in a voxel is normalized to the number $N_s$ of simulated events. In the post-processing of (Antunes et al., 2025), the resulting value was multiplied by $N_D$ calculated according to Eq. (10). Therefore, for a dose of 2 Gy, the dose value given in the last row of Table 3 were effectively multiplied by the ratio $N_{2\,Gy}/N_s$ given in the last column of Table 2. The contribution of the voxel to the mean dose in the nucleus and the mean square of the dose according to Eq. (3) are determined by a further division by the number of voxels in the nucleus (third row of Table 3).

The resulting estimated contributions of a single voxel containing an AuNP subject to ionizing interaction to the mean dose and mean square of the dose in the nucleus are listed in Table 4 for the four combinations of cell models and irradiation geometry. The contribution to the mean dose is two orders of magnitude smaller than the considered dose value of 2 Gy. On the other hand, the contribution to the square of the dose is between 9 Gy² and 36 Gy², comparable to the estimate of 20 Gy² determined at the end of Section 3.1.

Table 3: Nucleus and voxel volume and number of voxels in the nucleus for the two cell types (Antunes et al., 2025), radius of a sphere of the same volume as the voxels according to Eq. (11) in the text, and corresponding mean dose for an ionization in a GNP centered in the sphere for the dose distribution shown in Fig. 6.

| Cell model | suspension | adherent |
|---|---|---|
| nucleus volume / μm³ | 420 | 281 |
| Voxel volume / μm³ | 2.65×10⁻² | 2.07×10⁻² |
| Voxels per nucleus | 1.59×10⁴ | 1.365×10⁴ |
| Sphere radius / nm | 185 | 170 |
| Mean dose in sphere / Gy | 14.7 | 17.1 |

Table 4: Estimated contributions to the mean dose and the mean square of the dose in a voxel from a single GNP undergoing an ionizing interaction.

| cell model | irradiation | Contribution to $\overline{D}$ / Gy | Contribution to $\overline{D^2}$ / Gy² |
|---|---|---|---|
| suspension | *z*-axis | 0.027 | 11.5 |
| suspension | *x*-axis | 0.028 | 12.6 |
| adherent | *z*-axis | 0.052 | 36.0 |
| adherent | *x*-axis | 0.026 | 9.2 |

## 4. Discussion

Several notable observations can be made about the results reported by (Antunes et al., 2025): The predicted survival probabilities for cells without AuNPs (colored symbols) fall on straight lines in the semilogarithmic plots shown in Fig. 3. The survival predicted from the mean dose to the nucleus is in some cases higher for irradiation with AuNPs than without AuNPs (long dashed lines in Fig. 4). The reduction in predicted survival according to the LEM compared to that based on the mean dose appears to occur in discrete steps (dot-dashed lines in Fig. 4) as a function of the proportion of AuNPs found in the nucleus. The predicted survival depends on the direction of irradiation (Fig. 1(b)).

Another notable observation is that the volume of the cell in suspension is by a factor of 1.5 larger than that of the adherent cell. This means that with the same number of AuNPs in the cell, the mass fraction in the cell is higher by a factor of 1.5 in the adherent cell, which leads to a bias when comparing dose enhancement due to AuNPs in the cells. Since there may be a distribution of cell volume for both populations, conclusions on differences between the two cell types (in suspension or adherent) should be based on results from an ensemble of cells randomly selected from the two populations.

Several details of the methodology and the results obtained are not described in (Antunes et al., 2025). For instance, the number of particles in the PSF and their energy spectrum are not ´given, and it is not clear whether the source in the second simulation was placed so that its center coincided with that of the cell, or whether it was centered with the nucleus as in (Antunes et al., 2024).

Some of the missing information, such as the absolute number of electrons leaving the AuNPs in the simulations, could be estimated from the data presented by (Antunes et al., 2025). In other cases, this was not possible, and plausibility considerations had to be made and data from other sources used. For example, the measured energy spectrum of another irradiation facility was used as a surrogate for the radiation field.

*4.1 Survival prediction without AuNPs*

It is not plausible that the survival predictions, shown as colored symbols in Fig. 3, should lie on straight lines, as such trends correspond to a pure exponential dependence on dose. This indicates that an error occurred in the calculation of these data, e.g. that the square in the second term in the exponent was omitted in the evaluation of Eq. (1) or that the quadratic term was neglected altogether.

Under this assumption, the dose values corresponding to the simulation results were calculated from the slope of the dashed lines in Fig. 3 and the value of $\alpha = 0.35$ Gy$^{-1}$ given in (Antunes et al., 2025). The ratio between this dose value and the corresponding value on the $x$-axis is 1.30 for the cell in suspension. For the adherent cell this ratio is 1.27 and 1.39 for irradiation in the $z$-direction and the $x$-direction, respectively.

Using the value of $\beta = 0.03$ Gy$^{-2}$ given by (Antunes et al., 2025), the survival curves corresponding to the simulation results were determined according to the LQ model. The resulting curves are shown as dot-dashed lines in Fig. 3. These curves fail to reproduce the survival curve obtained from the experiments and underestimate the survival of the cells. This means that the agreement between the simulation results and experimental observations, indicated by the colored symbols in Fig. 3 falling into the gray shaded areas of the estimated error bands of the fit to the experimental data, is the result of a mistake.

It can be assumed that this disparity reflects the different irradiation conditions in the simulations compared to the experiments and that the number of photons hitting the cells was overestimated. This conjecture will be further explored in the second part of the paper.

*4.2 Impact of the dose gradient at the surface*

The results shown in Fig. 3 and Fig. 4 suggest that cells containing AuNPs receive a lower dose than cells without AuNPs when they are in suspension and irradiated along the $x$-axis or when they are adherent and irradiated along the $z$-axis. This dose reduction is in the order of several tens of percent.

Such a reduction of the mean dose due to the presence of AuNPs is not to be expected and is not physically plausible, even if the incident radiation field only consisted of the $^{60}$Co gamma rays. The mass-energy absorption coefficient of gold for 1.33 MeV photons (higher-energy $^{60}$Co gamma line) is a few percent lower than that of water. However, the linear energy absorption and attenuation coefficients at this energy are by more than a factor 17 higher for gold than for water. Therefore, there are more photon interactions and resulting energy deposition with AuNPs than without. This is all the more true for photons with lower energy, which are contained in the photon spectrum shown in Supplementary Fig. 2, in which the main gamma lines account for only about half of the fluence.

In Fig. 5, the dose shows an initial increase with depth at the surface. This means that the dose to a cell near the surface is significantly lower than near the dose maximum, where charged particle equilibrium prevails. Therefore, the irradiation conditions differ to which cells are exposed in an irradiation experiment, and the results from simulation and experiment are not expected to match. In addition, the variation with depth entails a potential source of bias, as the mean dose to a cell or its nucleus varies depending on the distance between the surface and the center of gravity of the cell or nucleus.

The estimated dose reduction for the floating cell from Fig. 4(b) was about a factor of 0.8, which is approximately equal to the ratio between the distance between the source and the center of the nucleus and between the source and the center of the cell, assuming that the irradiation in Fig. 2(a) was from the left side. A dose reduction of about 0.8 was also estimated for the adherent cell and irradiation along the $x$-axis, and a similar ratio between the two source distances is also found in Table 1 when the irradiation was from the left side in Fig. 2(b). For irradiation along the $z$-axis, the distances between the source and the centers of adherent cell and its nucleus show a ratio of about 0.8 which roughly agrees with the estimated dose reduction of about 0.7. The only case in which the rough estimates of the ratio between the two distances to the source do not agree with the dose reduction is the irradiation of the floating cell along the $z$-axis.

With the caveat that these are rough estimates, it can be stated that there is evidence that the physically implausible increase in cell survival when irradiating with AuNPs is most likely an artifact caused by two factors: that the simulation setup implies a strong dose gradient and that the mean dose to the cell was used for cells without AuNPs and the mean dose to the nucleus for cells with AuNPs.

## 4.3 Survival prediction with AuNPs

In general, it can be observed that the survival predictions shown in Fig. 4 tend to lie close to one of the three horizontal lines shown. Clear differences between the prediction from the mean dose to the nucleus and the LEM can only be seen in Fig. 4(a) for the cell in suspension with 100 % of AuNPs in the nucleus and in Fig. 4(b) for both cell models with 100 % NPs in the nucleus and for the adherent cell also at 50% and 75% AuNPs in the nucleus. Furthermore, the reduced survival values appear to be almost identical. This suggests that the simulations are biased from insufficient statistics of the data in the PSF.

The contributions to the mean dose and the mean square dose from a single event in the complete simulation in which an ionizing photon interaction occurs in an AuNP were estimated in Section 3.4. The results listed in Table 4 show that such a singular event does not significantly change the mean dose in the nucleus. However, the contribution to the mean square of the dose in the nucleus is in the order of 20 Gy$^2$, i.e. the value corresponding to the step changes in the survival prediction in Fig. 3.

That the lower values of predicted cell survival only occurred in some of the simulations is to be expected when the probability of such an event occurring is as small as one in $10^7$. However, with such a low probability, the observed changes are not statistically significant and do not allow any sound conclusions to be drawn.

The analysis of the energy spectrum of the emitted electrons (Supplementary Fig. 3) in Section 3.3 led to an upper limit for estimated number of ionizing photon interactions of 7, which is likely an overestimation of the actual number of such events in a simulation like that of (Antunes et al., 2025). In fact, Table 2 shows that the expected number of events (out of $10^7$) in which a photon hits an AuNP is in the order of 3000, while the expected number of events in which an ionizing interaction occurs in an AuNP is in the order of $4\times10^{-3}$. This means, an ionization in an AuNP by a photon occurs on average in about 1 out of 400 such simulations. This makes the results reported by Antunes *et al* (2025) very unlikely from a statistical point of view and raises the question of why differences were found between simulations with and without AuNPs.

Answering this question requires a detailed analysis of the arrangement of the AuNP positions in the cells and of the content of the PSF used as a source in the simulations. The use of PSFs is a suitable surrogate for the determination of particle fluence, which offers the benefit that the information obtained in the first simulation can be reused for several subsequent simulations. However, the phase-space coordinates sampled by a PSF always have discrete values and lead to bias if the initial positions of the photons and the AuNP positions coincide by chance.

This bias cannot be removed by repeating the simulation several times, unless the geometry is changed between the different repetitions. The placement of the NP must have been altered between the different cases, which represent different proportions of the NPs in the cell nucleus. However, it is not clear from the work of (Antunes et al., 2025) whether the placement of AuNPs varied between different simulations of the same scenario. If this was not the case and matches between the positions of photons and NPs occurred, the same bias persisted in all simulations of the same scenario.

The work of (Antunes et al., 2025) does not include information on the number of events in the PSF, the spatial distribution of the particles, their type and their energy spectrum. Such information would be necessary to substantiate the results obtained. If the number of entries in the PSF was smaller than $10^7$, geometrical coincidences would be amplified by repeated scoring of the same entries.

If the PSF contained mainly electrons, the estimated total number of particles hitting an AuNP in Table 2 would relate to electrons rather than photons and would be consistent with the number of electrons estimated in Section 3.3. These would be electrons traversing the AuNP rather than being generated in it. However, such a dominance of electrons in the PSF is also physically implausible and could only occur if there was a severe problem with the first simulation in which the PSF was produced.

The fact that events with ionizing photon interactions were found in the simulation of (Antunes et al., 2025) despite the low probability could be explained if one assumes that an error occurred and the AuNP diameter was 48.9 nm instead of 4.89 nm. In this case, the expected number of interactions was higher by a factor of 1000, so that the estimates for the number of events with ionizing photon interaction in an AuNP were in the order of 4. Then the observed results were plausible in the sense of understandable. They were still compromised by the fact that they were based on sparse events and far from providing a meaningful estimate of what would be observed in an experiment with many cells.

## 5. Conclusions

The analysis performed here indicates that the results reported by (Antunes et al., 2025) for $^{60}$Co irradiation are affected by methodological problems. The agreement between simulations and experimental data for cells without AuNPs seems to be due to a mistake in the calculation of the survival probability, probably neglecting the quadratic term of the linear-quadratic survival model. The simulation setup relates to cells near the surface, where the strong dose gradient causes variations in geometry to result in different doses to the cell and the nucleus.

The step-like changes between the survival predicted from the mean dose and the LEM suggest that only one event in the entire simulation involved an ionizing interaction of a photon in an AuNP. It was shown above that the probability of this is in the parts per thousand range. The total number of electrons leaving an AuNP, estimated from the reported electron spectrum, is three orders of magnitude higher than the value estimated from the expectation of photon interactions in the AuNPs.

This contradiction would be resolved if the AuNP diameter in the simulations were a factor of 10 larger than intended. Another possible explanation for the discrepancies is a hidden bias in the simulation geometry, e.g. if the distribution of AuNPs was non-uniform or coincidences between AuNP and photon start positions occurred.

In summary, the results reported by (Antunes et al., 2025) for $^{60}$Co irradiation appear to be affected by methodological problems and do not allow conclusions about the magnitude of effects of cell geometry on dose enhancement and radio-sensitization by AuNPs.

Many of the issues raised in this part of the paper could be addressed by running a larger number of simulations using a slightly different setup that ensures that the center of gravity of different scoring volumes is at the same distance from the surface. In addition, the positions of nanoparticles would have to be resampled each time a simulation is repeated, and the lateral positions of particles retrieved from the PSF should be homogenized. However, some aspects of the simulation setup, namely the considerations of cells near the surface and the use of the LEM comparatively large voxels, lead to systematic effects and bias that cannot be eliminated by increasing the number of simulations. The quantification of this bias is the subject of the second part of the paper.

## Acknowledgment

Oswald Mkanda acknowledges a scholarship of the German Academic Exchange Service (DAAD). Leo Thomas is acknowledged for performing the simulations that provided the data on energy deposition around a gold nanoparticle.

# Supplementary Figures and Tables

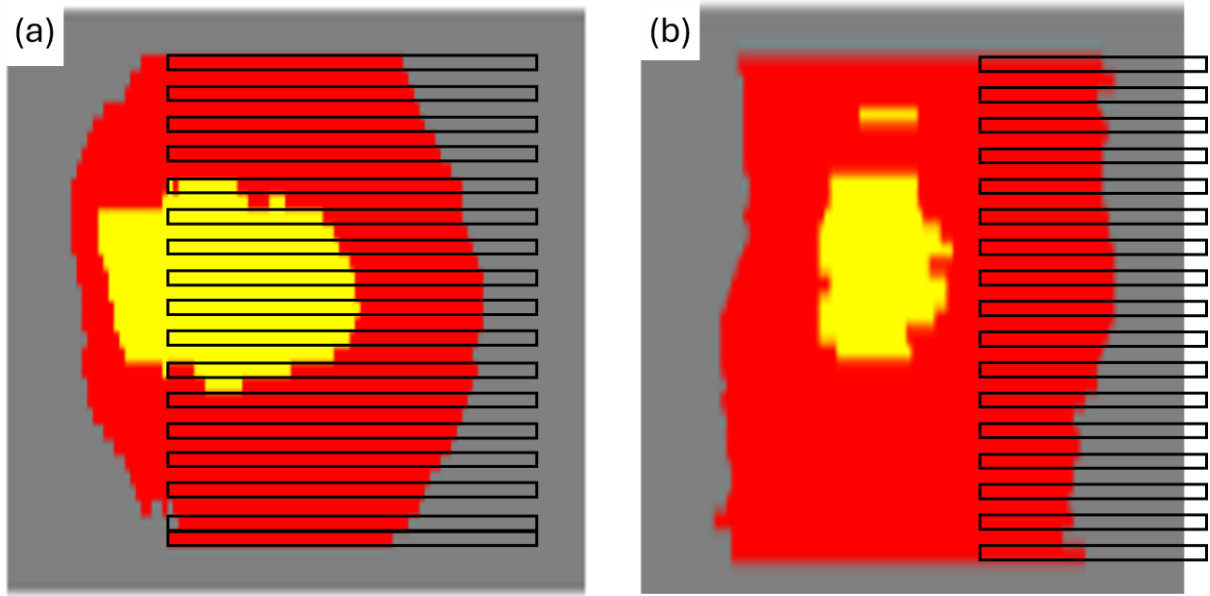

Supplementary Fig. 1: Cross-sections in the *y-z*-plane of the models for (a) a cell in suspension and (b) an adherent cell. Reproduced from (Antunes et al., 2025) (copyright Antunes et al.) under the CC BY 4.0 license (http://creativecommons.org/licenses/by/4.0/) with the following modifications: The panels were cut out, scaled along both directions, collated, and the labels and black rectangles were added.

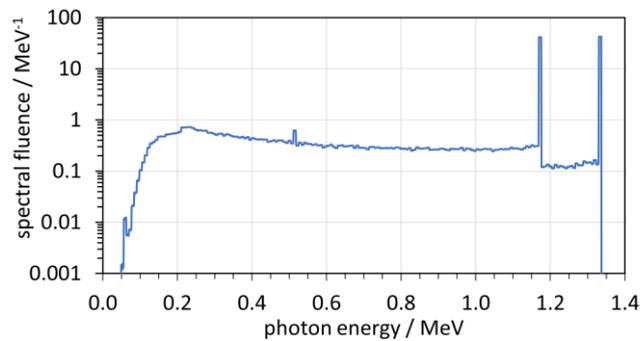

Supplementary Fig. 2: Photon energy spectrum at the PTB $^{60}$Co calibration facility at 1 m distance from the source (10 cm×10 cm reference field).

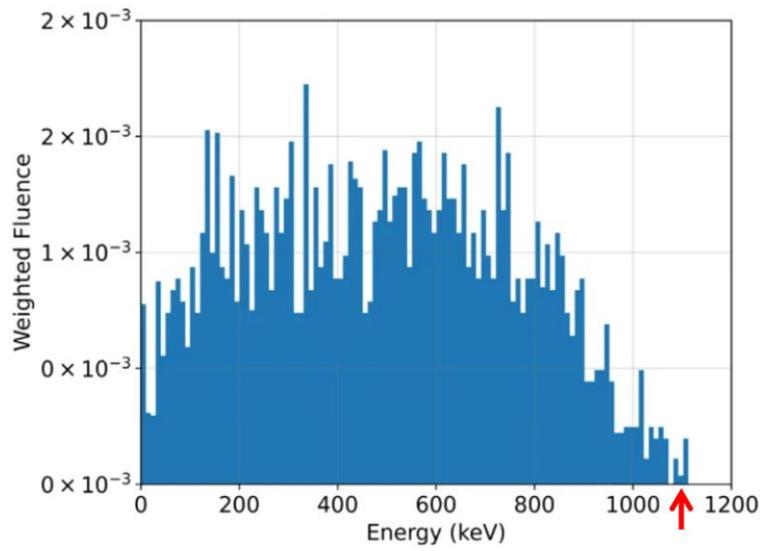

Supplementary Fig. 3: Relative fluence spectrum of electron leaving one of the 4326 spherical gold nanoparticles of 4.79 nm diameter. The arrow indicates the lowest fluence value in the spectrum. Reproduced from (Antunes et al., 2025) (copyright Antunes et al.) under the CC BY 4.0 license (http://creativecommons.org/licenses/by/4.0/), modified by cropping to the panel shown, removing the label "A", and adding the red arrow on the x-axis.